\documentclass[final,english]{bullsrsl}

\usepackage[latin1]{inputenc}

\usepackage[T1]{fontenc}

\usepackage{natbib} 

\usepackage{graphicx}
\usepackage{amsmath}
\usepackage{amssymb}
\usepackage{hyperref}
\hypersetup{colorlinks=true, linkcolor=blue, filecolor=blue, urlcolor=blue,citecolor=blue}

\begin{document}
\title{Effect of Multi-Species Plasma on Fanaroff-Riley Radio Jets}



  \author[affil={1,2}, corresponding]{Priyesh Kumar}{Tripathi}
  \author[affil={1}]{Indranil}{Chattopadhyay}
  \author[affil={3}]{Raj Kishor}{Joshi}
  \author[affil={4}]{Ritaban}{Chatterjee}
  \author[affil={1}]{Sanjit}{Debnath}
  \author[affil={2}]{M. Saleem}{Khan}

\affiliation[1]{Aryabhatta Research Institute of Observational Sciences (ARIES), Manora Peak, Nainital, 263001, India}
\affiliation[2]{Department of Applied Physics, Mahatma Jyotiba Phule Rohilkhand University, Bareilly, 243006, India}
\affiliation[3]{Nicolaus Copernicus Astronomical Center, Polish Academy of Sciences, Bartycka 18, PL-00-716 Warsaw, Poland}
\affiliation[4]{School of Astrophysics, Presidency University, 86/1 College Street, Kolkata, West Bengal, India 700073}


\correspondance{priyeshkumarankur@gmail.com, priyesh@aries.res.in}


\maketitle

\begin{abstract}

The Fanaroff-Riley (FR) dichotomy observed in extragalactic radio jets has been attributed to a range of possible mechanisms, including intrinsic jet properties such as the presence of different species in the plasma. Jet material may span from a pure electron-positron pair plasma to mixed plasmas containing electrons, positrons, and protons, or even to hadronic jets made up of electrons and protons only. To investigate this aspect, we present results from three-dimensional simulations of low-power, supersonic, magnetized jets at kiloparsec scales in a magnetohydrodynamic framework. By varying the plasma composition, we show its impact on jet stability and on the development of diffuse structures typical of core-brightened FR type I sources. Our results indicate that the growth of non-axisymmetric instabilities plays a key role in disrupting the jet head.
\end{abstract}





\section{\label{sec:Intro}Introduction}
Back in 1974, \citeauthor{1974_Fanaroff_Riley_MNRAS} introduced a broad classification of extended extragalactic radio jets into two distinct types by correlating their radio morphology with luminosity. Low-power sources, classified as Fanaroff-Riley type I (FR I), show core-brightened morphologies with faint, diffuse outer lobes, whereas high-power sources, classified as Fanaroff-Riley type II (FR II), show edge-brightened lobes with prominent hotspots located far from the central nucleus. This dichotomy has been interpreted through several factors, including differences in jet power supplied by the central engine \citep{1992_Baum_ApJ}, and variation in the properties of their ambient environments \citep{1993_DeYoung_ApJ,1995_Bicknell_ApJS,2002_Gopal-Krishna_Wiita_NewAR,2014_Perucho_et_al}. In addition, the intrinsic properties of the jet itself, particularly its particle composition (dominated by electron-positron pairs or electron-proton plasma or a mixture of electron, positron, and proton), may also contribute in shaping these observed divisions in FR classes, \citep{1996_Reynolds_MNRAS,1997_Celotti_MNRAS,2018_Croston_MNRAS}.
There are several numerical simulations that have significantly contributed to advancing our understanding of this FR dichotomy and have successfully reproduced the jet morphologies \citep{2007_Perucho_Marti,2016_Tchekhovskoy_Bromberg_MNRAS,2016_Massaglia_et_al_A&A,2019_Massaglia_et_al_A&A,2020_Rossi_et_al_A&A}. However, detailed information of the plasma composition is often absent, with only a few exceptions \citep{scheck02,2023_Joshi_Chattopadhyay_ApJ...948...13J}. However, these simulations were 2-dimensional, which suppressed some of the instabilities and therefore showed quantitative differences in the jet morphology due to the difference in the composition of the jet. 
To explore this aspect, we conduct large-scale three-dimensional magnetohydrodynamic (3D-MHD) simulations of low-power, supersonic magnetized jets that have already been decelerated to non-relativistic speeds on kiloparsec scales. We simulate jets with very different plasma compositions to assess whether distinct morphological behavior can emerge from their underlying composition. We study the subsequent jet stability and the transition toward diffuse and turbulent flow structures in these jets. 

The outline of the paper is as follows. We describe the MHD equations used in our simulations, along with the initial setup in Section~\ref{sec:Eqn and Setup}. Then, the results are presented in Section~\ref{sec:Results}, followed by the summary in Section~\ref{sec:Summary}.

\section{\label{sec:Eqn and Setup}Governing Equations of MHD and Simulation Setup}
   The time-dependent ideal magnetohydrodynamics (MHD) equation of motion is written in the conservative form as:
        \begin{equation}
     \label{eq:MHD eqns}
         \frac{\partial}{\partial t} 
         \begin{pmatrix} \rho \\ \rho\vec{v} \\ E \\ \vec{B} \\ \rho\Phi \end{pmatrix} + 
         \vec{\nabla} \cdot
         \begin{pmatrix} \rho\vec{v} \\  \rho{\vec{v} \otimes \vec{v}}-{\vec{B} \otimes \vec{B}}+p^*\vec{I} \\ (E+p^*){\vec{v}-\vec{B}(\vec{v}\cdot \vec{B})} \\ {\vec{v} \otimes \vec{B} - \vec{B} \otimes \vec{v} } \\ \rho\vec{v}\Phi \end{pmatrix} = 0   
     \end{equation} 

  where $\rho, \vec{v}$, and $\vec{B}$ are the total mass density, bulk velocity, and magnetic field, respectively, $p^*$ is the sum of thermal pressure and magnetic pressure, and $E$ is the sum of thermal, kinetic, and magnetic energy density of the plasma. $\Phi$ represents a tracer field that is used to distinguish between the pure and diffused jet material.
  For the closure of the MHD equations \eqref{eq:MHD eqns}, we use relativistically correct \citet{2009Chattopadhyay&RyuApJ} equation of state (CR EoS) for multispecies flows. The CR EoS can account for the effects of multispecies plasma through a composition parameter $\xi$, which is the ratio of proton number density to electron number density. That means $\xi=1$ corresponds to electron-proton plasma, $\xi=0$ corresponds to electron-positron pair plasma, and in between values of $\xi$ would be a mixture of electron-positron-proton plasma. CR EoS is widely used in literature for various studies \citep{2011_Chattopadhyay_Chakrabarti_IJMPD..20.1597C,2022_Joshi_etal_ApJ...933...75J,2023_Sarkar_etal_MNRAS.522.3735S,2024_Joshi_et_al_ApJ...971...13J,2024_Debnath_et_al_MNRAS.528.3964D,2025Debnath_et_al,2025Tripathi_et_al_979_61}.
  We solve the conservative set of equations \eqref{eq:MHD eqns} using our in-house-developed multi-dimensional, Godunov-type, finite-volume MHD simulation code. The details of the simulation code are given in \citet{2025Debnath_et_al,2026Tripathi_et_al}.

\begin{figure}
\centering
\includegraphics[width=\textwidth]{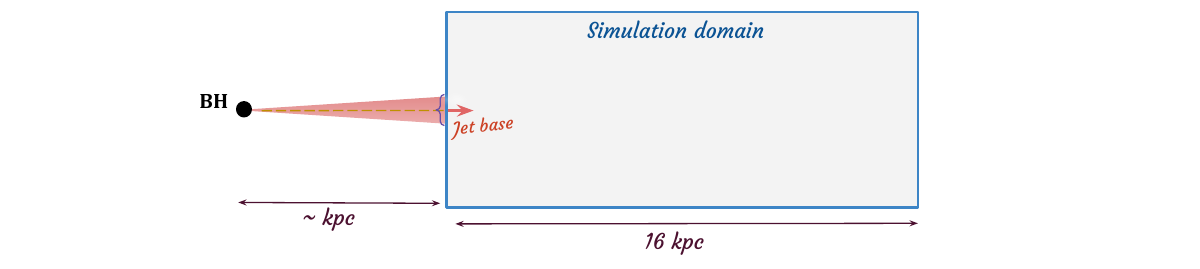}
\begin{minipage}{12cm}
\centering
\caption{Schematic diagram of simulation setup.} \label{fig:Cartoon}
\end{minipage}
\end{figure}
  
  The simulations are performed in a three-dimensional Cartesian domain, with the $z$-direction corresponding to the jet's propagation direction. An illustration of our simulation setup is shown in Figure~\ref{fig:Cartoon}, showing the jet base located at the center of the lower $z$-boundary. We inject the jet beam through this base, which is situated at kiloparsec scales from the central black hole. Since our focus is not on modeling the initial jet deceleration, which occurs between the launching region and the kiloparsec distances \citep{1999_Laing_et_al_MNRAS,2014_Laing_Bridle_MNRAS}. Instead, we assume that by the time the jet beam enters the simulation domain at these kiloparsec scales, it has already been decelerated to non-relativistic velocities by some means. 
  We chose the radius of the jet beam ($r_{\rm j}=100$ parsec) and the speed of light ($c$) as the units of length and velocity, respectively. The simulation domain extends to $160~r_{\rm j}$ (corresponding to 16 kiloparsec) in the $z$-direction, with a uniform resolution of six grid cells per beam radius. The ambient medium is initially taken to be unmagnetized, stationary, and follows a \citet{1972_King_ApJ} density profile. The supersonic jet beam is injected through a nozzle at the jet base, having a constant density, and an azimuthal (toroidal) magnetic field configuration as prescribed in \citet{1989_Lind_et_al_ApJ...344...89L}.
  This choice of toroidal magnetic field at injection is motivated by two reasons: (i) VLA polarimetric study of FR I jets \citep{2014_Laing_Bridle_MNRAS} has shown the transition of magnetic field from axial into toroidal configuration being dominant at larger distances, and (ii) the inclusion of the longitudinal ($B_z$) component will introduce additional acceleration in the jet along the z-direction. Since our focus is to study the transition of initially supersonic jets towards subsonic diffused structures typical of FR I morphology and the role of plasma composition in driving these transitions through current-driven kink instabilities, we consider only the toroidal part of the magnetic field in this work.
  For a detailed description of the setup, one may refer to \citet{2026Tripathi_et_al}.

\section{\label{sec:Results}Results}
\begin{figure}
\centering
\includegraphics[width=\textwidth]{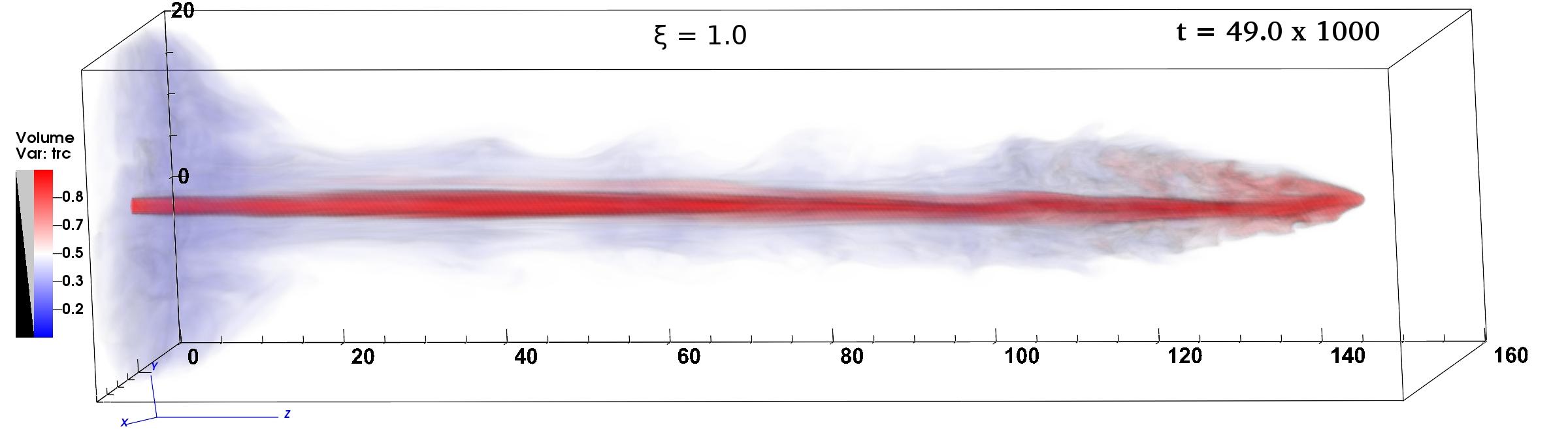}
\includegraphics[width=\textwidth]{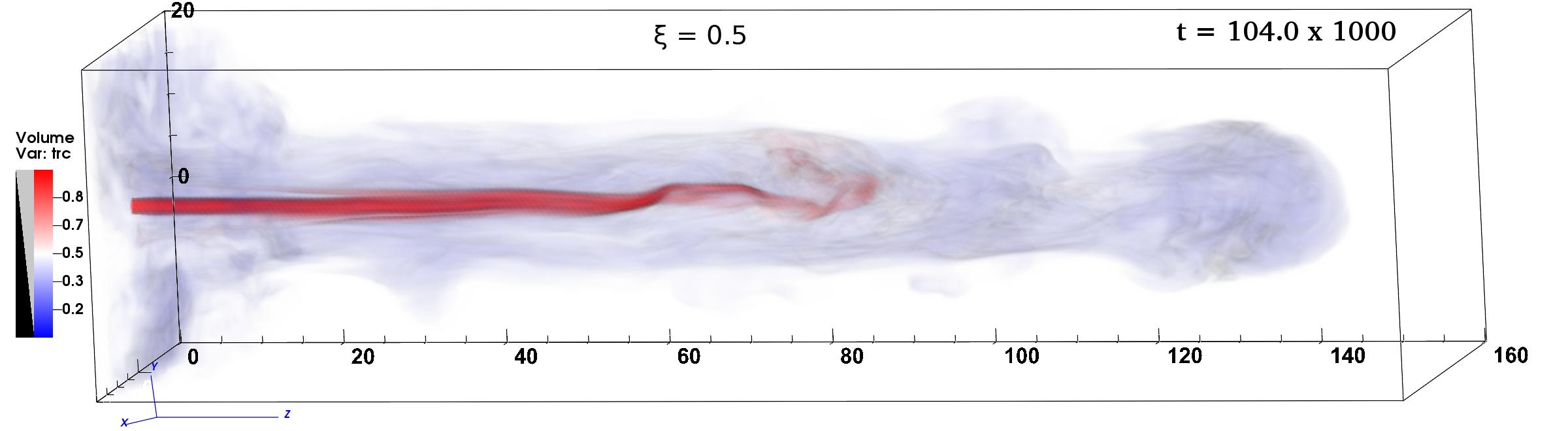}
\includegraphics[width=\textwidth]{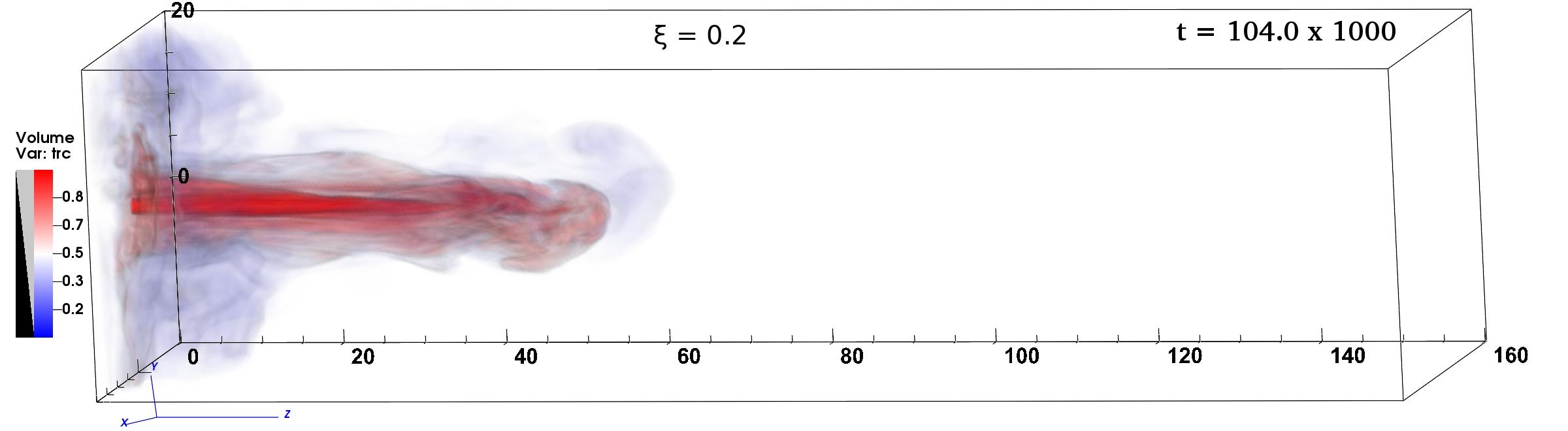}
\begin{minipage}{12cm}
\centering
\caption{3D volume rendering of the jet tracer for three models having different compositions ($\xi=1.0,0.5,0.2$) showing their morphology.} \label{fig:trcVolume}
\end{minipage}
\end{figure}

\begin{figure}
\centering
\includegraphics[width=\textwidth]{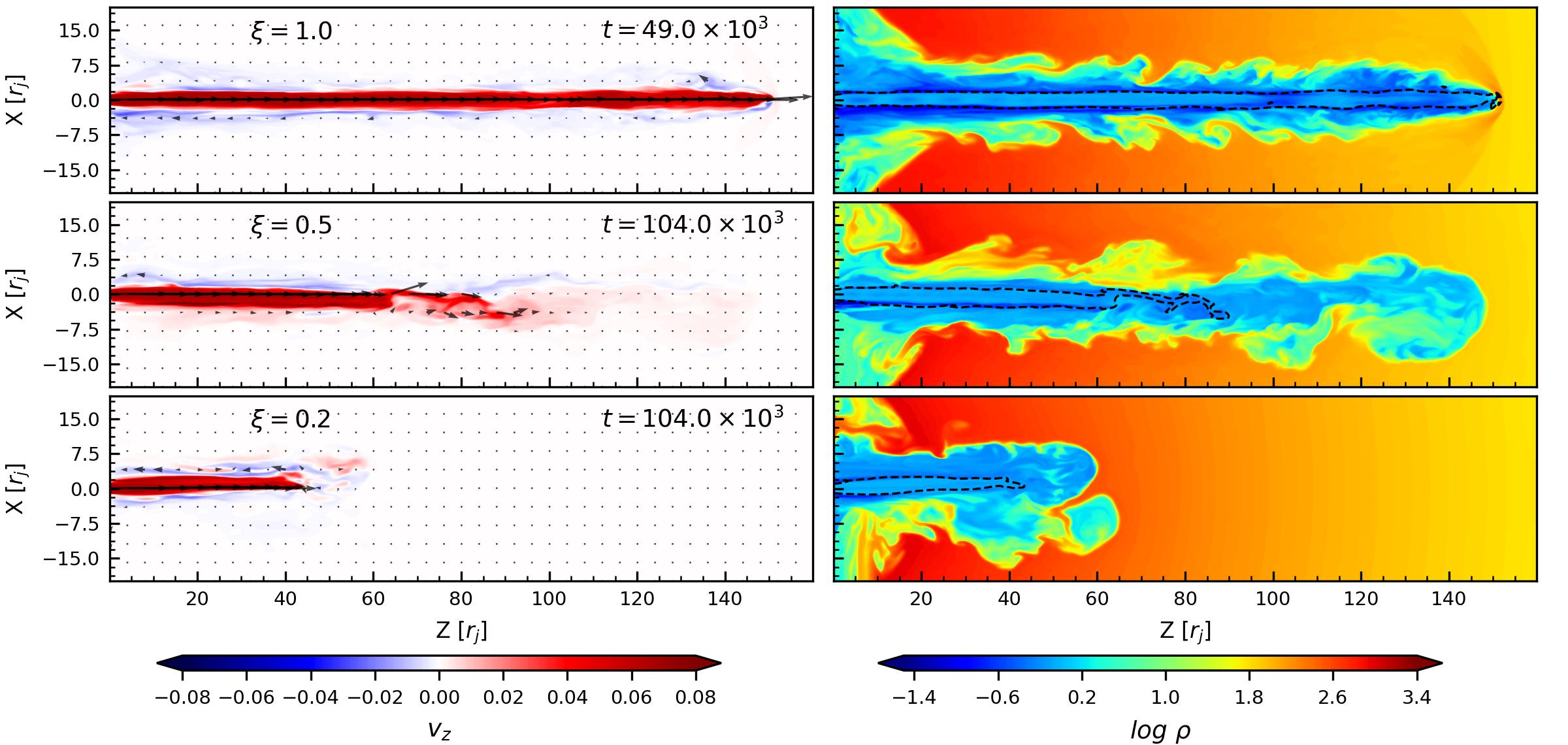}
\begin{minipage}{12cm}
\centering
\caption{$X$-$Z$ plane view for three models in different row panels. Left: axial velocity ($v_z$) contours, overlaid with velocity vectors (black arrows), Right: log density contours. The black dashed line shows the Mach-$1$ surface. } \label{fig:vzRho2D}
\end{minipage}
\end{figure}

   Figure~\ref{fig:trcVolume} compares the morphologies of three different jet models from our simulations that differ in their plasma compositions ($\xi=1.0,~0.5,~0.2$). The plot shows the three-dimensional volume rendering of the jet tracer distribution. The red regions correspond to the pure jet material with a tracer value of unity, while the white and blue regions indicate the diffused jet material mixed with the ambient medium. The length is expressed in units of the jet radius ($r_{\rm j}$), and time in units of $r_{\rm j}/c$. All three models are launched with identical velocity ($0.067$c) at the jet base, with similar temperatures, magnetic fields (plasma-$\beta=10$), and jet-to-ambient density contrasts ($10^{-3}$).
   This density contrast at injection is chosen to simulate a low-power jet with kinetic luminosity of ($\sim10^{42}$erg s$^{-1}$) for the given injected velocity.
   Figure~\ref{fig:vzRho2D} further shows the contours of the axial velocity ($v_z$) and logarithmic density in the left and right panels, respectively, for an $X$-$Z$ slice of each model in separate rows.  
   The first model ($\xi=1.0$), representing a jet made of pure electron-proton plasma, shows a well-collimated jet beam propagated up to $\sim 150r_{\rm j}$ as reflected by both tracer and axial velocity plots (top rows of Figure~\ref{fig:trcVolume} and~\ref{fig:vzRho2D}). This jet terminates at a forward shock clearly visible in density contours at the head of the jet (right panel of Figure~\ref{fig:vzRho2D}). This forward shock is responsible for the presence of a hot spot at the jet head, as observed in FR II sources \citep{2022_Massaglia_et_al_A&A}.

   The jet in the second model ($\xi=0.5$) is composed of multi-species plasmas with equal proportions of protons and positrons, maintaining overall charge neutrality with electrons. The enrichment of positrons makes the jet relatively hotter at the same physical temperature, causing a weakening in the Mach disk \citep{2002_Gopal-Krishna_Wiita_NewAR}. The hotter jet will now be more likely to move laterally. As a result, the jet becomes more susceptible to non-axisymmetric current-driven instabilities like kink modes \citep{1993_Eichler_ApJ,2000_Appl_el_al,2016BrombergTchekhovskoy}, which caused the jet beam to deviate from its straight course (see middle panel of Figure~\ref{fig:trcVolume} and~\ref{fig:vzRho2D}). The black dashed line in the right panel of Figure~\ref{fig:vzRho2D} marks the Mach-1 surface, indicating the transition between supersonic and subsonic flow. It can be seen that the jet head's propagation has also slowed to subsonic speeds. Consequently, the jet is disrupted, and the diffused jet material is produced at the head, as shown in the middle panel of Figure~\ref{fig:trcVolume}, which is no longer a collimated structure. This jet lacks a strong forward shock at the head because its energy gradually dissipates along the beam, making the beam appear relatively brighter than a faint jet head, similar to FR I sources.
    
   The bottom panel of Figure~\ref{fig:trcVolume} and~\ref{fig:vzRho2D} shows another multi-species plasma jet, where the protons and positrons have a ratio of $1:4$ ($\xi=0.2$). This additional enrichment of the positron fraction further reduces the strength of the Mach disk. The jet has a higher tendency to move in lateral directions such that it stalls beyond a certain distance. We can see in Figure~\ref{fig:trcVolume} and~\ref{fig:vzRho2D} that even at a similar time $t=104.0 \times 10^3$, the diffused material in the $\xi=0.5$ jet has propagated till $\sim 140r_{\rm j}$, but the $\xi=0.2$ jet is unable to propagate beyond $50r_{\rm j}$. Additionally, the forward shock is absent in this model too.

\begin{figure}
\centering
\includegraphics[width=0.8\textwidth]{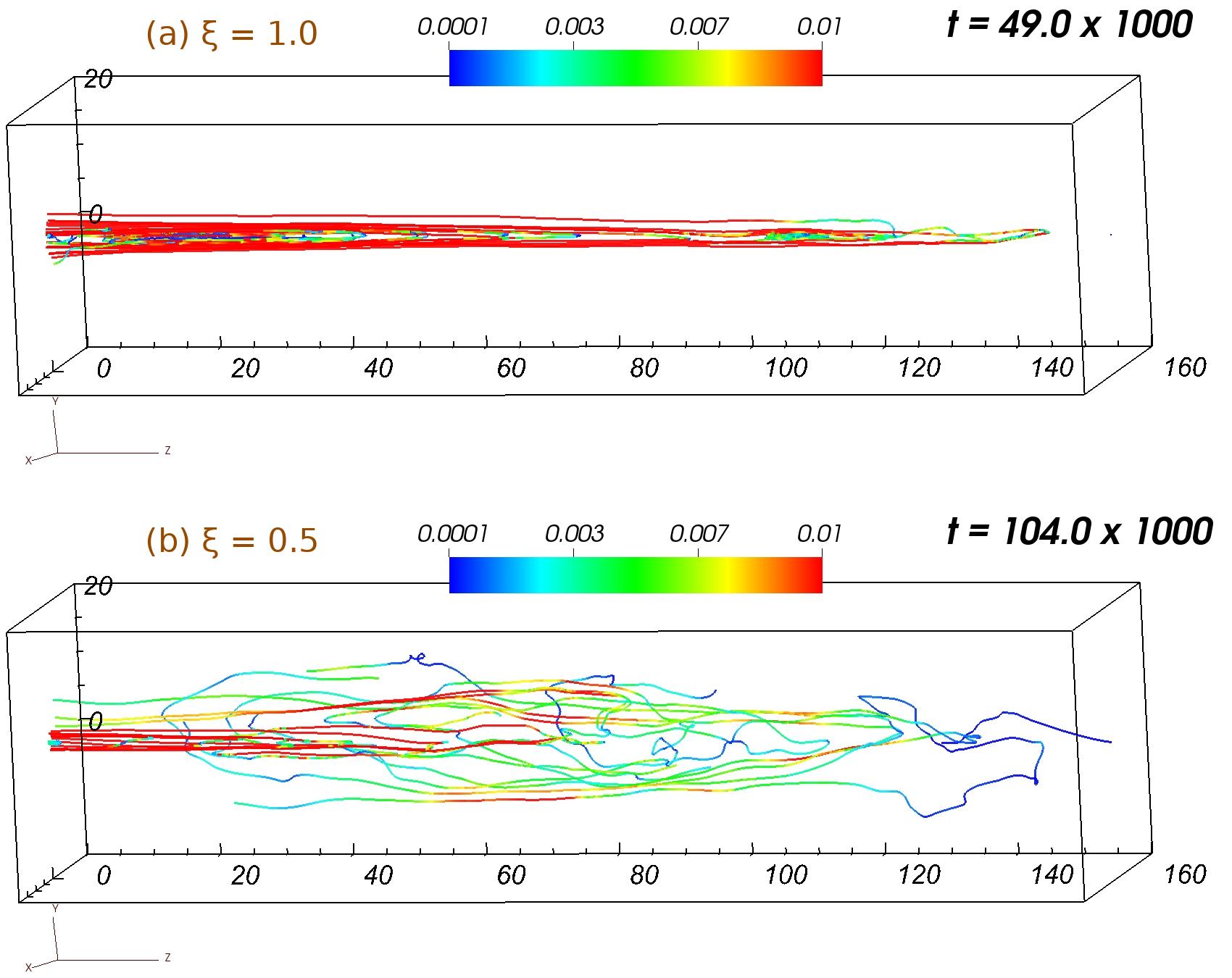}
\begin{minipage}{12cm}
\centering
\caption{3D view of streamlines of the magnetic field for $\xi=1.0~\&~0.5$ jets. The color depicts the strength of the magnetic field.} \label{fig:streamlines}
\end{minipage}
\end{figure}

   Figure~\ref{fig:streamlines} presents the three-dimensional magnetic field streamlines for $\xi=1.0~\&~0.5$ jet models. The magnetic field, which was purely azimuthal at the injection, has been stretched out by the jet, generating a significant axial component ($B_z$). In the electron-proton ($\xi=1.0$) jet, the streamlines remain largely coherent and are stretched almost to the jet head, producing stronger $B_z$ components. In contrast, the multi-species ($\xi=0.5$) jet displays a different magnetic topology. Ahead of the jet beam, at $z=60r_{\rm j}$, the magnetic field becomes highly distorted, with streamlines forming irregular structures. This is accompanied by significant spreading of the magnetic field lines in the lateral directions, indicating mixing within the flow. 
   
\begin{figure}
\centering
\includegraphics[width=\textwidth]{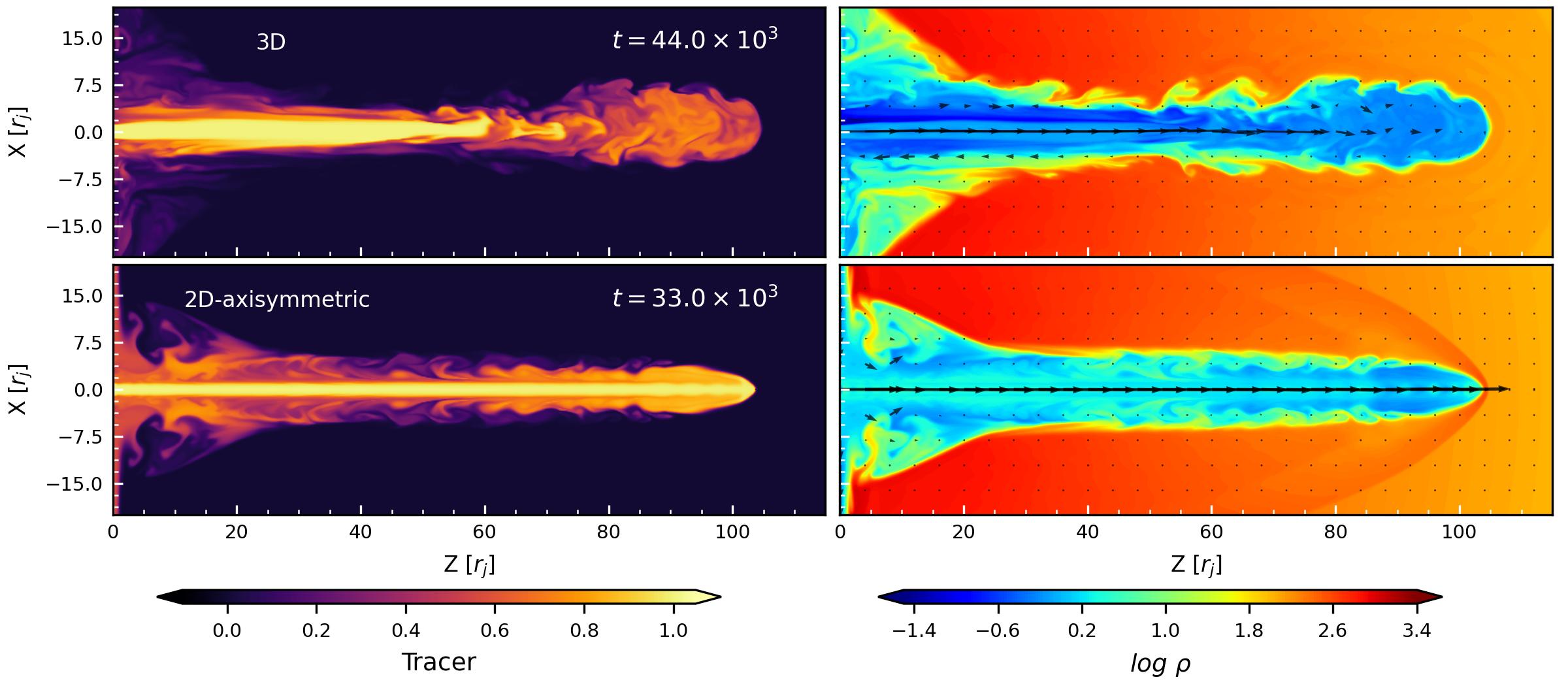}
\begin{minipage}{12cm}
\centering
\caption{Comparison of 3D jet simulation with 2D-axisymmetric simulation for same injection parameters. The left panel shows the jet tracer distributions, and the right panel shows the log density contours, overlaid with velocity vectors (black).} \label{fig:2Dvs3D}
\end{minipage}
\end{figure} 
   
   Figure~\ref{fig:2Dvs3D} compares the results of a 3D jet simulation with those of a 2D-axisymmetric one, where the initial injection parameters are identical in both cases. The jet is composed of electron-proton plasma ($\xi=1.0$), but has a higher physical temperature, leading to a diffuse jet head in 3D. The plots are tracer distributions on the left and logarithmic density on the right. One can see here the presence of diffusion in the jet head in the 3D simulation, whereas it is entirely absent in the 2D-axisymmetric case. The disparity arises from the additional degree of freedom available in the 3D case, which allows the jet to dissipate energy through small-scale, non-axisymmetric motions. In low-power jets, this enhanced energy dissipation prevents the forward shock from being sustained, eventually leading to its disappearance. In contrast, the imposed condition of axisymmetry in the 2D case artificially preserves the forward shock, producing an over-collimated "nose cone" structure at the jet head \citep{2010_Mignone_et_al_MNRAS.402....7M}. This feature is a well-known artifact of axisymmetric simulations, arising from the suppression of non-axisymmetric current-driven instabilities in jets.
   That is why the 3D simulations are necessary to accurately capture the disruption and transition of the jet towards diffuse morphologies \citep{2016_Massaglia_et_al_A&A} typical of core-brightened FR I sources.

\section{\label{sec:Summary}Summary and Discussions}
   In this work, we have presented results from three simulations with different plasma compositions to address the role of multi-species plasma in the FR dichotomy observed in radio-loud extragalactic jets. Using a magnetohydrodynamic framework with a relativistic equation of state, we simulated low-power jets propagating at kiloparsec scales. We found that the morphology of these jets depends on plasma composition, primarily because of differences in their thermodynamic properties. Electron-proton jets show strong forward shocks and develop FR II-like features, whereas jets enriched with positrons become thermally hotter and increasingly susceptible to non-axisymmetric instabilities.
   The growth of these non-axisymmetric kink instabilities plays a major role in deforming and disrupting the jet head, allowing the formation of diffuse structures at large distances.

   It may be noted that we took a modest resolution (6 cells to resolve the beam radius at injection) in this study. A higher resolution would have resolved the internal structures of the jet beam better, but the large-scale evolution and overall morphology would remain qualitatively unchanged. The kink instability seen here is a large-scale bending of the jet beam from its axis that disrupts the jet spine when the kink mode's growth timescale is shorter than the advection timescale in the jet. If it had been an effect of numerical dissipation due to low resolution, it would have appeared in all the models rather than following a pattern. In this study, kink mode growth depends on the Mach disk being weaker, which results from a change in plasma composition even for flows injected with the same temperature.
   Moreover, recent studies based on large samples from Low-Frequency Array (LOFAR) observations have revealed significant overlap in radio power between FR I and FR II populations, presenting several outliers, such as low-power FR IIs and high-power FR Is \citep{2019Mingo_et_al}. These findings suggest that radio luminosity alone does not reliably predict morphology.


\begin{acknowledgments}
We thank the anonymous referee for providing constructive comments, which significantly improved the quality of the paper. The authors are grateful to the Indian and Belgian funding agencies DST (DST/INT/Belg/P09/2017) and BELSPO (BL/33/IN12) for granting financial support to organize the fourth BINA workshop and other BINA activities.
PKT acknowledges the use of computational resources provided by ARIES \texttt{Surya} and IUCAA \texttt{Pegasus} HPC clusters for conducting the 3D simulations presented in this work. RC thanks ANRF for a SURE grant (SUR/2022/001503) and IUCAA for their hospitality and usage of their facilities during his stay at different times as part of the university associateship program.
\end{acknowledgments}

\begin{furtherinformation}

\begin{orcids}

  \orcid{0009-0002-7498-6899}{Priyesh Kumar}{Tripathi}
  \orcid{0000-0002-2133-9324}{Indranil}{Chattopadhyay}
  \orcid{0000-0002-9036-681X}{Raj Kishor}{Joshi}
  \orcid{0000-0001-9899-7686}{Ritaban}{Chatterjee}
  \orcid{0000-0002-9851-8064}{Sanjit}{Debnath}
\end{orcids}

\begin{authorcontributions}
PKT has run the simulations presented here. IC supervised the project. All co-authors have contributed to the writing and reviewing of this manuscript.
\end{authorcontributions}

\begin{conflictsofinterest}
The authors declare that there is no conflict of interest.
\end{conflictsofinterest}

\end{furtherinformation}



%

\bibliographystyle{bullsrsl-en}

\bibliography{extra}

\end{document}